# Well-being in Isolation: Exploring Artistic Immersive Virtual Environments in a Simulated Lunar Habitat to Alleviate Asthenia Symptoms

Grzegorz Pochwatko*
Institute of Psychology
Polish Academy of Sciences

Wieslaw Kopec
Polish-Japanese Academy
of Information Technology

Justyna Swidrak
Institute of Psychology - PAS
Fundació de Recerca Clínic
Barcelona - IDIBAPS

Anna Jaskulska
Kobo Association

Kinga H. Skorupska
Polish-Japanese Academy
of Information Technology

Barbara Karpowicz
Polish-Japanese Academy
of Information Technology

Rafał Masłyk
Polish-Japanese Academy
of Information Technology

Maciej Grzeszczuk
Polish-Japanese Academy
of Information Technology

Steven Barnes
Emotion Cognition Lab
SWPS University

Paulina Borkiewicz
Visual Narratives Laboratory
Lodz Film School

Paweł Kobyliński
National Information
Processing Institute

Michał Pabiś-Orzeszyna
Institute of Contemporary
Culture University of Lodz

Robert Balas
Institute of Psychology
Polish Academy of Sciences

Jagoda Lazarek
Polish-Japanese Academy
of Information Technology

Florian Dufresne
Arts et Métiers
Institute of Technology

Leonie Bensch
Software for Space Systems
and Interactive Visualization
German Aerospace Center

Tommy Nilsson
European Space Agency (ESA)

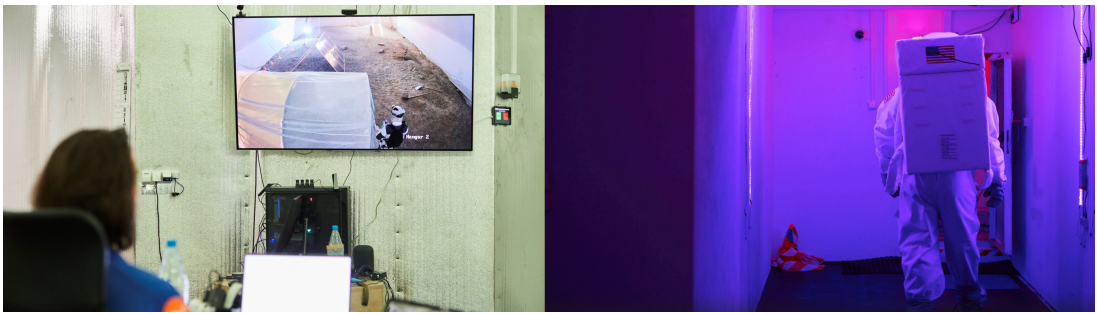

Figure 1: Left: Crew remotely supporting the EVA team during their simulated moonwalk. Right: Entering the EVA area.

**ABSTRACT**

Revived interest in lunar and planetary exploration is heralding a new era for human spaceflight, characterized by frequent strain on astronaut's mental well-being, which stems from increased exposure to isolated, confined, and extreme (ICE) conditions. Whilst Immersive Virtual Reality (IVR) has been employed to facilitate self-help interventions to mitigate challenges caused by isolated environments in several domains, its applicability in support of future space expeditions remains largely unexplored. To address this limitation, we administered the use of distinct IVR environments to crew members (n=5) partaking in a simulated lunar habitat study. Utilizing a Bayesian approach to scrutinize small group data, we discovered a significant relationship between IVR usage and a reduction in perceived stress-related symptoms, particularly those associated with asthenia (syndrome often linked to chronic fatigue and weakness; a condition characterized by feelings of energy depletion or exhaustion that can be amplified in ICE conditions). The reductions were most prominent with the use of interactive virtual environments. The 'Aesthetic Realities' - virtual environments conceived as art exhibits - received exceptional praise from our participants. These environments mark a fascinating convergence of art and science, holding promise to mitigate effects related to isolation in spaceflight training and beyond.

## 1 RATIONALE

Spearheaded by NASA's Artemis program, ongoing efforts to reestablish human presence on the Moon face challenges to astronaut well-being resulting from prolonged periods of cohabitation in ICE (isolated, confined and extreme) conditions of the lunar environment [11]. Numerous relevant psychoenvironmental factors, such as congestion, lack of privacy, social isolation, or sensory deprivation [35, 41], have been identified as potential triggers of behavioral problems, including anger, anxiety, interpersonal conflict, sleep disorders, and decreased motivation, all of which can increase the risk of mission failure [14, 18, 26]. Such factors have prompted the European Space Agency (ESA) to acknowledge the significance of exploring these psychological challenges and identifying appropriate countermeasures in its road-map for future space research [1].

Mitigating the adverse impacts of ICE is a key focus of both actual space missions as well as simulations conducted in terrestrial analogue environments and experimental deployments in artificial test-bed facilities, especially long-duration ones [5]. Among the recognized means are:

a) providing access to professional help (e.g. the opportunity to consult with personnel trained in the detection of early symptoms of disorders - either a trained crew member or ground personnel),

---

*e-mail: gp@psych.pan.pl

b) allowing individuals to bring their favorite food and some personal items onboard,
c) facilitating personal communication with family and friends on Earth,
d) providing internet access to maintain connectivity,
e) allocating free time for leisure activities e.g. reading, playing games or engaging with art forms (e.g. NASA astronaut Chris Hadfield performed the song Space Oddity onboard the ISS).

We may hypothesize that the utilization of Virtual Reality (VR) presents an additional approach to alleviate the impacts of ICE in forthcoming missions. Yet, there are challenges to using VR in space, especially related to payload weight and power consumption. However, standalone headsets largely address them. In fact, a modified Oculus Quest 1 was already tested in space, as part of the 2021-22 "Cosmic Kiss" mission to the ISS where it served to facilitate on-board training [15]. Hence, the use of VR is feasible in space, so it could be used help mitigate ICE conditions by allowing users to experience social contexts (alleviating isolation), vast landscapes (alleviating confinement) and providing a break from the routine and responsibilities (alleviating the extreme nature of space habitation). VR applications have a rich history in various pertinent fields, spanning from mental and physical health applications [6, 13, 13, 21], to the realm of military training simulations [33, 49]. These applications hold promise in providing adaptable, cost-effective, and location independent immersive simulation of remote or fictional environments. For these reasons, VR is enjoying growing popularity among space agencies [16]. In the mental health domain, one of the most recognized uses is for the treatment of phobias (e.g., fear of heights, agoraphobia, fear of public speaking, fear of flying, and arachnophobia) and to reduce symptoms of PTSD for soldiers or car accident survivors [6]. Importantly, future space missions seldom require the treatment of mental diseases, but primarily the prevention of the former to ensure the safety and performance of the astronauts throughout the mission. However, especially the usage of VR as a potential tool for the prevention of mental health issues still remains underexplored in research. Here, only a few dedicated studies can be found [2, 52]. In particular, none of these studies is related to the use of art as a form of prevention or treatment. Recently, it has been shown that the combination of art and VR alleviate anxiety and enhance affective states [39]. However, the benefits of using art as a means of maintaining or improving well-being during long space expeditions have already been recognized several times in the space context [34, 45, 46].

Building upon these insights, our study has two objectives:

a) To investigate whether artistic VR experiences can enhance the emotional and physical well-being of participants in ICE conditions.
b) To compare the impact of cinematic ART VR and interactive ART VR on well-being, evaluating their effectiveness in this regard.

## 2 METHODS

### 2.1 Simulated Isolation Mission: Habitat, Participants, and Schedule

In view of long-duration missions to the lunar surface, space agencies around the world have increasingly invested in preparatory activities, in particular Earth-based analogue missions [9]. Typically, such activities have taken place in remote or extreme natural environments that approximate some of the challenges associated with living on the Moon. The ESA-operated Concordia station on Antarctica, for instance, has been used as a test-bed for simulation of future lunar expeditions [22]. By utilizing Earth-based analog facilities that replicate ICE conditions relevant to human space exploration, valuable insights into the challenges faced by future astronauts have been gained. These insights have typically been derived from studying a relatively limited number of participants, yet they have proven to be significant in helping researchers understand the effects of ICE conditions [23].

Following this line of reasoning, the analog facility used in this study provides a fully enveloping environment that mimics ICE conditions in a lunar habitat, including complete isolation. The crew consisted of five volunteers recruited from among individuals affiliated with the research institutions involved. They were two women and three men in their twenties or thirties, while one crew member was 40+. They belonged to two nationalities and communicated in English. Three of them had solid prior VR experience, one had some VR experience and one none at all. Their roles were divided into Commander, First Officer, Chief Scientist, Chief Engineer, and Medical Officer. The crew was supported by a mission control team and research control team via regular voice briefings. During this 14-day mission, the crew only left the habitat for extravehicular activities (EVA) in an enclosed dome that simulates the lunar surface, as seen in Fig. 1.

### 2.2 The Interactive and Cinematic Artistic VR Environments.

Within the context of ICE training, artistic VR, where the viewers become active participants in crafting theirs artistic experiences [27], could serve as a distinctive method to enhance participants' emotional resilience and overall well-being. By crafting immersive escapes from simulation conditions and sparking creativity and positive emotions, artistic VR may not only facilitate stress reduction but also augment participants' sense of agency, potentially alleviating feelings of confinement and isolation inherent in such training scenarios. As an artistic VR environment, we chose the Nightsss experience in two parallel versions: Interactive VR [30] and Cinematic VR [29]. The content of both versions is almost identical, the only difference is the fact that in the Interactive VR version the user has the ability to move freely through the 3D environment and influence its state (e.g. by touching objects and observing their reactions), while in the Cinematic VR version the user becomes a passive spectator of a 360 3D movie. Nightsss made its debut at the Sundance festival and has been shown in both forms at subsequent festivals and shows in many countries. According to its creators, the Nightsss is a virtual erotic poem with elements of artistic animation and ASMR (Fig. 2, see video figure [3]). It is based on a piece by the spoken word poet W. Lewandowska. In a virtual night environment, the users meet a dancing character, which from time to time turns into other shapes, comes and goes. While the character 'dances' the poem, the rhythmic transformation leads from night to a daytime environment that brings a meditative space for reflection on one's body and movement in virtual space. Users see their hands in the created world that serves as an anchor. The choice of the Nightsss environment was dictated by methodological considerations: it is the only available artistic VE appearing in two versions identical in terms of content, differing only in form. This allows comparisons to be made between the effects of interactivity and the number of degrees of freedom on the dependent variables.

### 2.3 Nature VR, the control environment.

The application of nature-themed VR experiences as a well-being improvement tool is supported by a growing body of research. It has been shown that exposure to natural environments, even in virtual form, can effectively reduce stress, improve mood, and enhance overall well-being [2, 51]. For instance, VR nature experiences have been used in healthcare settings to alleviate anxiety and improve patient outcomes [19, 20]. Moreover, the immersive nature of VR can potentially amplify the restorative effects of nature exposure, providing an escape from confined and stressful environments, a feature especially valuable in contexts like space missions. Given the isolation and confinement, Nature VR experiences could offer an invaluable respite and contribute to maintaining psychological health.

In Nature VR participants could admire the natural landscape from an observation terrace on a hill, mapped onto the physical 4x4 m space, in which the users could freely walk around. The scene was arranged to create the impression of open space, but at the same time limit it in a discreet way. This was to convince users that they could explore, but also make them stay in the mapped virtual space. This allows us to avoid breaks in presence [37]. The virtual experience was accompanied by sounds of nature. The duration was 7 minutes to match the duration of the artistic experience. The environment was adapted from other studies in which its effectiveness as a neutral control condition was verified [36].

### 2.4 Research Tools

The virtual environments were displayed on the HTC Vive Pro Eye HMD with wireless add-on and an Intel I9 2.8 GHz PC featuring a Gigabyte Geforce RTX 3070 graphics card. Heart rate and electrodermal activity (EDA) were measured with the use of Fitbit Sense smartwatches. Furthermore, we have used the adapted [48] and shortened version of the Singapore Mental Wellbeing Scale (SMWEB-S) [17], the Pictorial Self-Assessment Mannequin (SAM) questionnaire to measure the emotional response [8, 28], Self-assessment of primary and secondary symptoms of asthenia converted into a survey based on the most common signs reported in the context of participants in space missions [42] as well as evaluation of virtual experiences (multidimensional assessment based on the questionnaire methods described below).

Scales being used:

a) multi-dimensional feedback tool designed by authors to evaluate participants' responses to presented virtual experiences. The tool employs a seven-point scale, gauging dimensions like positivity/negativity, calmness/excitement, and desire for future participation.
b) Short Presence Scale (SPS) created by authors, based on the original MEC-SPQ questionnaire [50] (e.g., I felt as if I were taking part in events, not just watching them)
c) The Self-Assessment Manikin (SAM) is a widely trusted, picture-based tool assessing emotional responses in valence (positive/negative), arousal (calm/excited), and dominance (controlled/controlling) areas. Its simplicity and broad use make it valuable for research in various fields, including technology interaction studies. [8]
d) short well-being scale adopted by authors - selected items from Mental Well-being Scale [17], e.g. "I am optimistic about the future", I am calm").
e) primary and secondary symptoms of asthenia - created by authors on the basis of [43] - rating on a 0- "I don't experience at all", 100- "I experience them to the full extent" scale (e.g. fatigue, sleep disturbance, weakness, memory disorder)

### 2.5 Procedure

#### 2.5.1 Mission Schedule and Daily Routine

The mission was conducted in the aforementioned lunar research facility [31]. Throughout the course of the mission, the participants engaged in a total of 5 EVA sessions. Their daily schedules were packed with various tasks, including EVA preparations and conducting experiments, as seen in Fig. 1. Each evening, participants filled out individual mission diaries, short momentary assessment surveys, and daily psychological questionnaires. Here, a self-assessment of the primary and secondary symptoms of asthenia was implemented.

#### 2.5.2 Main manipulation - Virtual experience days

On the eighth, tenth, and twelfth day of the simulation, virtual experiences were introduced. Each participant was exposed to one of the VR experiences on each of these days. The order of experiments for each participant was randomly selected before the start of the mission and was not known beforehand. Only the technical officer learned about the types of experiences intended for individual participants (to prepare and implement them). On the day of the virtual experience, participants took baseline EDA measurements. This was always done around 8.30 am. Then the participants started to carry out the mission activities planned for the day, such as collecting bio-samples, engineering EVA solutions, planning and conducting EVA missions, and managing and maintaining resources. Virtual experiences were implemented in the early afternoon, individually. The timing of the experiences was directed by the strict mission schedule, as the participants followed a repetitive routine. Participants measured pre-VR EDA with their Fitbit watches and filled out questionnaires on their personal tablets (SAM and well-being). Then, with the help of a technical officer, they put on a VR headset and trackers and participated in one of the virtual experiences. After completing the virtual experience, they remeasured EDA (post-VR measure) and completed questionnaires (presence, SAM and well-being).

#### 2.5.3 Interviews

Individual interviews with each mission participant were conducted shortly after leaving the habitat. To ensure maximum comfort, the participants could choose their preferred location for the interview - they all opted for an open-air setting in the vicinity of the habitat. The interviews followed a flexible structure with some fixed elements. The first element involved the participants sharing their spontaneous thoughts. Specifically, they were asked to list the first thoughts that come to their mind after completing the mission. They were then prompted to talk about the most important things for them that they remembered. They were asked to reflect on both positive and negative experiences related to the activities conducted during the mission, as well as their interpersonal relationships. Additionally, they were asked to talk about the VR experiences and to express their opinions regarding the use of virtual environments in future missions.

## 3 Results

### 3.1 Data preparation

The mission participants completed the questionnaires and recorded their data using personal tablets. Questionnaire responses were recorded in the cloud. Data from Fitbit Sense watches were stored in the Fitbit app and cloud. In order to maintain anonymity, the accounts of individual participants were encrypted, no personal e-mails were used. Once the data was downloaded, it was permanently deleted from the cloud. Skin conductance data (single recordings) were visually inspected and corrected for artifacts (e.g. sudden breaks in signal - see [4, 7]). Upon inspection, it was revealed that while complete data was collected for all other data types, only the physiological data from participant 2 and 5 could be subjected to analysis. The files for the remaining astronauts were either incomplete or contained substantial amounts of missing data. Notably, the data from A2 and A5 have been reported since they were collected without any missing values or significant artifacts. In contrast, other analog astronauts encountered challenges with physiological data collection due to the absence of external communication, resulting in partial physiological data availability for A1, A3, and A4. Comprehensive qualitative data was successfully gathered from all analog astronauts and analyzed accordingly. It's important to acknowledge that these statistics are derived from a notably small sample size, rendering them preliminary and exploratory in nature. Their primary purpose is to pave the way for more extensive replication studies. To address this limitation, we have employed a Bayesian analysis of data from a single-case approach.

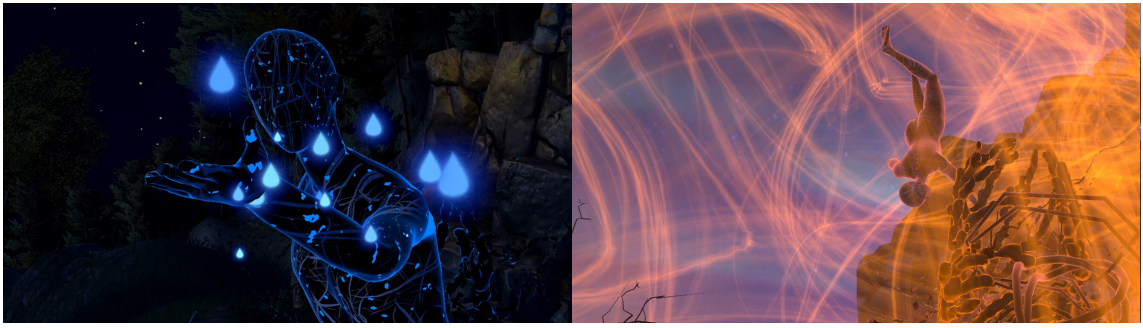

Figure 2: Frames from the "Nightsss/Noccc" experience, dir. by Weronika Lewandowska, Sandra Frydrysiak; night (left) and day (right).

The problems encountered indicate the need to refine the procedure for measuring psychophysiological variables with the use of wearables. Nevertheless, the obtained accurate data is adequate for conducting a case study analysis. Due to limited data resolution (1 Hz recording), we opted to analyze the overall mean skin conductance level (SCL). Averaged values from two-minute intervals for baseline, pre- and post-VR sessions were computed. Separate means are presented for each virtual experience type.

Processing of raw data and analysis of VR's impact on asthenia symptoms were conducted using the R software, v.4.2.0 Vigorous Calisthenics [38]. We utilized the JASP software for the evaluation of other models [24].

#### 3.1.1 Bayesian approach to single case and small sample data analysis

We use Bayesian approach to analyse small group data. It gives opportunity to analyse small samples or even single cases (e.g. treatment results in extremely rare diseases; novel drug testing [10, 40]). In highly specialized research fields, it can often be challenging to recruit a large sample. Our study focuses on a very specific population, limiting the potential pool of participants. A Bayesian mixed-effects model can be beneficial in such contexts. It offers a valuable resolution to the analytical challenges posed by our unique dataset. While small sample sizes may limit the statistical power and generalizability of our findings, they do not preclude valuable insights. The results derived from this study should be viewed as a stepping-stone for further research, providing preliminary evidence in a less explored area. We acknowledge that our findings may benefit from validation with larger samples, and caution should be applied when interpreting the results. Still, they significantly contribute to comprehending a distinct population within unique circumstances, illuminating complex variable interactions within a highly specific context.

Table 1: Evaluation of the VR experience on a scale from boring to interesting (differences from intercept)

| Level | Estimate | SE | 95% CI Lower | 95% CI Upper |
|---|---|---|---|---|
| Cinematic VR | −0.589 | 0.438 | −1.505 | 0.334 |
| Interactive VR | 1.980 | 0.433 | 1.043 | 2.887 |
| Nature VR | −1.391 | 0.430 | −2.299 | −0.493 |

Intercept=4.007; SE=0.455; 95% CI [2.957, 5.061]; R-hat 1.000; ESS (bulk)2776.703; ESS (tail) 2845.118

### 3.2 Evaluation of virtual experiences.

Factors determining the effectiveness of an intervention in the form of artistic VR may be related to its positive evaluation and acceptance. Artistic Interactive VR was perceived as the most interesting and liked. All participants assessed it highly. The Cinematic VR ratings were very scattered, but on average clearly lower than those of the interactive version. It was also the least liked experience. Nature VR was rated as uninteresting and of average liking (see Fig. 3). Bayesian linear mixed models were applied. In the "boring-interesting," dimension the baseline level was estimated at 4.007. Relative to this baseline, Cinematic VR showed no significant difference (Estimate: -0.589, CI: -1.505 to 0.334), Interactive VR showed a significant increase (Estimate: 1.980, CI: 1.043 to 2.887), and Nature VR indicated a significant decrease (Estimate: -1.391, CI: -2.299 to -0.493); see Table 1. With regards to "liking" the VR experience, the baseline level was 4.334. Compared to this, Cinematic VR significantly decreased (Estimate: -1.733, CI: -2.774 to -0.673), Interactive VR significantly increased (Estimate: 1.855, CI: 0.791 to 2.913), and Nature VR showed no significant difference (Estimate: -0.122, CI: -1.227 to 0.966); see Table 3. All participants expressed a desire to engage in similar artistic interactive experiences again. The remaining two experiences have mixed results (see Fig. 3). The baseline level was 4.303. Relative to this, Cinematic VR decreased but was not significantly different (Estimate: -1.322, CI: -2.746 to 0.078), Interactive VR significantly increased (Estimate: 1.648, CI: 0.252 to 2.972), and Nature VR showed no significant difference (Estimate: -0.326, CI: -1.753 to 1.130); see Table 4.

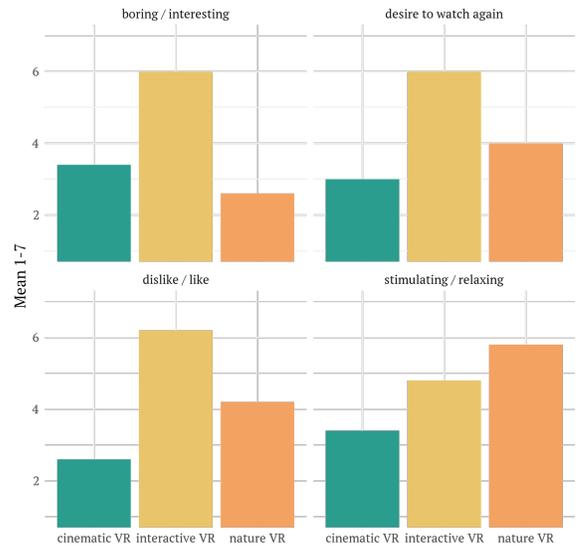

Figure 3: Perception of VR experience and desire to watch similar experiences again.

Table 2: Evaluation of the VR experience in terms of its stimulating versus relaxing qualities (differences from intercept)

| Level | Estimate | SE | 95% CI Lower | 95% CI Upper |
|---|---|---|---|---|
| Cinematic VR | −1.265 | 0.505 | −2.349 | −0.171 |
| Interactive VR | 0.140 | 0.517 | −0.959 | 1.246 |
| Nature VR | 1.125 | 0.507 | 0.033 | 2.188 |

Intercept=4.686; SE=0.576; 95% CI [3.337, 6.086]; R-hat 1.002; ESS (bulk)1762.301; ESS (tail) 1224.549

Table 3: Liking of the VR experience (differences from intercept)

| Level | Estimate | SE | 95% CI Lower | 95% CI Upper |
|---|---|---|---|---|
| Cinematic VR | −1.733 | 0.487 | −2.774 | −0.673 |
| Interactive VR | 1.855 | 0.502 | 0.791 | 2.913 |
| Nature VR | −0.122 | 0.493 | −1.227 | 0.966 |

Intercept=4.334; SE=0.515; 95% CI [3.072, 5.525]; R-hat 1.001; ESS (bulk)2606.871; ESS (tail) 2361.095

The above results are consistent with the commentaries obtained during interviews, in which the Interactive VR was described as the most interesting, Nature VR as quite boring, and Cinematic VR as the worst of all in many respects, including comfort, lack of interactivity, inability to move in the presented world, and discomfort when attempting movement (head-locked spheres moved with the participant causing break in presence).

### 3.3 Emotions and well-being

The change in declared well-being (measured with short well-being scale) is marginal for all types of virtual experiences (see Table 5. Further research is needed to explore this topic more comprehensively, as insignificant differences prevent us from drawing definitive conclusions about whether we have averted a decline in well-being or if there is no effect of manipulation.

Additionally, the Cinematic VR experience resulted in a notable decrease in positive emotions after the VR exposure compared to the pre-experience baseline. The estimate of the difference in happiness level was -0.923 with a 95% confidence interval ranging from -1.583 to -0.263, indicating a substantial and significant deviation from the pre-experience happiness level. Both Interactive VR and Nature VR environments showed only minor changes in happiness levels from the pre- to the post-VR experience. For the Interactive VR experience, there was a minor increase in happiness post-VR (estimate: 0.465), but the 95% confidence interval (-0.230 to 1.176) overlapped with zero, suggesting this change was not statistically significant. The Nature VR experience, similarly, showed minimal change in happiness levels from pre to post-experience, with an estimate of 0.072 and a 95% confidence interval (-0.600 to 0.757) encompassing zero (see Table 6). Model convergence seemed good as indicated by R-hat values close to 1. The Effective Sample Size (ESS) for both bulk and tail was satisfactory, suggesting adequate sampling for the posterior distributions. However, wide confidence intervals signify uncertainty in these estimates.

There were no significant differences between VR environments in the levels of self reported arousal and control(see Fig. 4, Table 8 and Table 7). Although the pattern of results is consistent with the indicators of psychophysiological arousal described below, it is important to approach these results with caution and recognize the need for further studies to obtain more precise estimates.

### 3.4 Physiological arousal

The results derived from EDA measurements should be treated as preliminary due to previously described equipment issues. Full

Table 4: Desire to participate in the VR experience again (differences from intercept)

| Level | Estimate | SE | 95% CI Lower | 95% CI Upper |
|---|---|---|---|---|
| Cinematic VR | −1.322 | 0.675 | −2.746 | 0.078 |
| Interactive VR | 1.648 | 0.642 | 0.252 | 2.972 |
| Nature VR | −0.326 | 0.658 | −1.753 | 1.130 |

Intercept=4.303; SE=0.677; 95% CI [2.672, 5.926]; R-hat 1.001; ESS (bulk)2095.652; ESS (tail) 2100.892

Table 5: Well-being scale; VR experience * pre-/post VR (differences from intercept)

| Level | Estimate | SE | 95% CI Lower | 95% CI Upper |
|---|---|---|---|---|
| Cinematic VR * pre | 0.021 | 0.112 | −0.208 | 0.263 |
| Interactive VR * pre | 0.121 | 0.119 | −0.133 | 0.371 |
| Nature VR * pre | −0.213 | 0.114 | −0.456 | 0.025 |
| Cinematic VR * post | 0.055 | 0.111 | −0.170 | 0.286 |
| Interactive VR * post | 0.025 | 0.119 | −0.220 | 0.270 |
| Nature VR * post | −0.010 | 0.117 | −0.263 | 0.242 |

Intercept=4.346; SE=0.177; 95% CI [3.959, 4.727]; R-hat 1.003; ESS (bulk)2240.964; ESS (tail) 3193.435

and analyzable data were only collected from two astronauts. For Astronaut 2, an increase in arousal was noticeable following the Interactive VR experience, while the other conditions led to significantly lower arousal (see Fig. 5 left). Interestingly, Cinematic VR did not cause notable differences in SCL compared to the baseline level immediately preceding the experience. The Nature VR condition led to a slight increase in arousal. Therefore, an increase in arousal can be observed in both interactive conditions for this participant. The arousal pattern of Astronaut 5 was similar in Nature and Cinematic VR conditions, but there was no clear increase in arousal for the Interactive VR condition (see Fig. 5 right). Due to the described data deficiencies, it is not possible to compare these results with the declarative scales (SAM calm/aroused and the experience rating as stimulating/relaxing).

### 3.5 Self-reported level of asthenia symptoms

We adopted a Bayesian statistical approach, utilizing the 'rjags' package in R. Analysis was focused on a Poisson log-linear model with four categories (pre VR, and post Nature, art Cinematic and Interactive VR). Furthermore, we computed differences between all possible pairs of these parameters, which provided deeper insight into their relative effects. The model's parameters were estimated using Markov chain Monte Carlo (MCMC) sampling. Convergence diagnostics were performed using the ESS.

The Bayesian approach has illuminated the effects of various forms of VR on the symptoms of asthenia. Our analysis demonstrates an overall trend towards decreasing symptom severity when using VR (see Table 9). This trend is captured by the mean values of the four VR experience levels, with 1 signifying pre VR and levels 2 to 4 representing Nature VR, art cinematic VR, and art interactive VR respectively. These average values decrease from 3.16 (pre VR) to 1.55 (Interactive VR). This negative trend suggests a VR related decrease in the perceived intensity of asthenia symptoms. The differences in symptom intensity between VR usage periods (diff.1-2, diff.1-3, etc.) emphasize the impact of different VR experiences. Both artistic VR environments appear to mitigate asthenia symptoms more effectively than the Nature control, with the interactive VR experience demonstrating the most significant decrease. The ESS values for each of the parameters (all above 2900) suggest a high

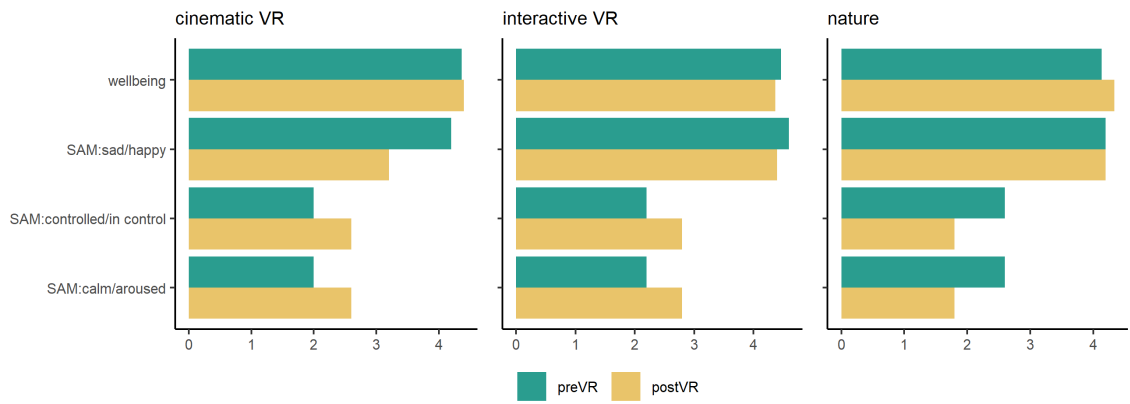

Figure 4: Well-being and SAM self report ratings by VE experience

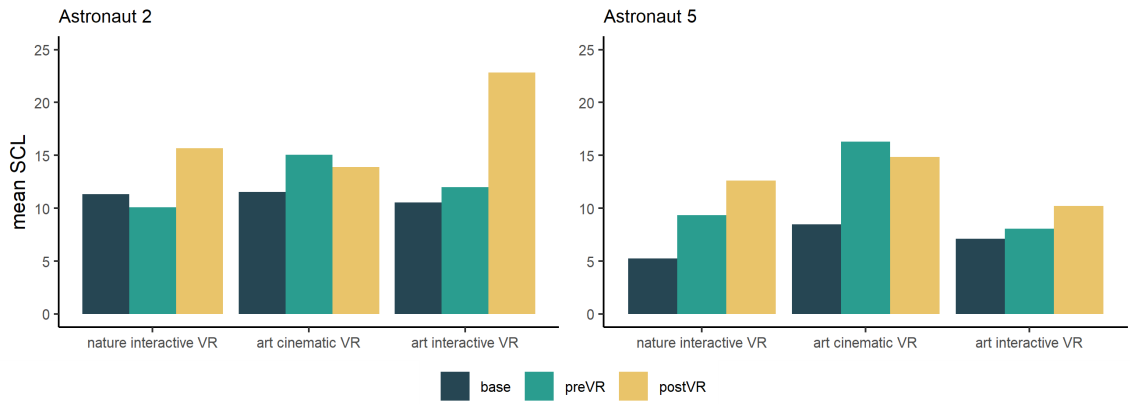

Figure 5: Mean skin conductance level [muS] (Astronauts 2 and 5)

Table 6: SAM (I am happy / sad); VR experience * pre-/post VR (differences from intercept)

| Level | Estimate | SE | 95% CI Lower | 95% CI Upper |
|---|---|---|---|---|
| Cinematic VR * pre | 0.053 | 0.326 | −0.605 | 0.703 |
| Interactive VR * pre | 0.465 | 0.341 | −0.230 | 1.176 |
| Nature VR * pre | 0.072 | 0.339 | −0.628 | 0.773 |
| Cinematic VR * post | −0.923 | 0.320 | −1.583 | −0.263 |
| Interactive VR * post | 0.261 | 0.344 | −0.427 | 0.947 |
| Nature VR * post | 0.072 | 0.332 | −0.600 | 0.757 |

Intercept=4.137; SE=0.210; 95% CI [3.659, 4.612]; R-hat 1.001; ESS (bulk)3229.167; ESS (tail) 3291.210

Table 7: SAM (I feel controlled / in control); VR experience * pre-/post VR (differences from intercept)

| Level | Estimate | SE | 95% CI Lower | 95% CI Upper |
|---|---|---|---|---|
| Cinematic VR * pre | 0.564 | 0.449 | −0.342 | 1.462 |
| Interactive VR * pre | −0.639 | 0.455 | −1.548 | 0.272 |
| Nature VR * pre | 0.370 | 0.447 | −0.535 | 1.291 |
| Cinematic VR * post | −0.831 | 0.457 | −1.748 | 0.111 |
| Interactive VR * post | −0.032 | 0.453 | −0.960 | 0.888 |
| Nature VR * post | 0.569 | 0.455 | −0.354 | 1.495 |

Intercept=3.836; SE=0.254; 95% CI [3.291, 4.378]; R-hat 1.001; ESS (bulk)5178.864; ESS (tail) 4062.850

degree of reliability and robustness in the analysis, as they are significantly larger than the common threshold of 1000, indicating that the MCMC chains have converged and provide a solid basis for the estimations. Therefore, these findings suggest that the application of VR, particularly the Interactive VR, can considerably alleviate asthenia symptoms.

### 3.6 Qualitative analyses

#### 3.6.1 Mission diaries

Among the open questions in the mission diaries, 21 entries were recorded relating to the reasons for the occurrence of asthenia symptoms, identified by the participants themselves. Upon analyzing them, several notable trends emerge. The most prevalent theme is the influence of interpersonal dynamics and conflicts within the crew. Instances of conflicts between crew members, both at an individual level and within the command structure, are cited multiple times. These conflicts seem to contribute to increased stress and irritability, which could exacerbate asthenia symptoms. Additionally, sleep deprivation consistently emerges as a contributing factor, aligning with existing research on the relationship between sleep quality and fatigue. Participants also frequently mention factors such as demanding tasks and extended extravehicular activities (EVA), which appear to impact their fatigue, subsequently intensifying asthenia manifestations. Other factors, such as emotional responses to mission

Table 8: SAM (I feel calm / in aroused); VR experience * prepost (differences from intercept)

| Level | Estimate | SE | 95% CI Lower | 95% CI Upper |
|---|---|---|---|---|
| Cinematic VR * 1 | −0.332 | 0.356 | −1.049 | 0.397 |
| Interactive VR * 1 | −0.129 | 0.348 | −0.874 | 0.638 |
| Nature VR * 1 | 0.270 | 0.388 | −0.534 | 1.074 |
| Cinematic VR * 2 | 0.263 | 0.347 | −0.483 | 0.998 |
| Interactive VR * 2 | 0.464 | 0.378 | −0.333 | 1.255 |
| natureVR * 2 | −0.535 | 0.362 | −1.283 | 0.232 |

Intercept=2.324; SE=0.312; 95% CI [1.573, 3.035]; R-hat 1.001; ESS (bulk)3275.446; ESS (tail) 3380.098

Table 9: VR experience and sympthoms of asthenia

| Variable | Mean | Q2.5 | Q50 | Q97.5 | ESS |
|---|---|---|---|---|---|
| pre VR [1] | 3.16 | 3.10 | 3.16 | 3.21 | 3133.67 |
| Nature VR [2] | 2.81 | 2.57 | 2.81 | 3.04 | 3144.31 |
| Cinematic VR [3] | 2.40 | 2.14 | 2.41 | 2.65 | 3002.97 |
| Interactive VR[4] | 1.55 | 0.85 | 1.57 | 2.13 | 2901.29 |
| diff.1-2 | -0.35 | -0.60 | -0.34 | -0.11 | 3162.14 |
| diff.1-3 | -0.75 | -1.03 | -0.75 | -0.50 | 3039.36 |
| diff.1-4 | -1.60 | -2.29 | -1.59 | -1.02 | 2900.24 |
| diff.2-3 | -0.41 | -0.76 | -0.40 | -0.05 | 3162.77 |
| diff.2-4 | -1.26 | -1.97 | -1.25 | -0.62 | 2964.28 |
| diff.3-4 | -0.85 | -1.57 | -0.84 | -0.21 | 2917.73 |

conclusion, diet deviations, and physical strain from exhaustive tasks, also contribute to the exacerbation of asthenia symptoms. Overall, the qualitative analysis underscores the intricate interplay between interpersonal interactions, sleep patterns, and mission-related challenges in influencing the severity of asthenia symptoms experienced by astronauts. Addressing these multifaceted aspects may hold the key to developing more effective strategies for mitigating asthenia and enhancing the psychological well-being of astronauts in isolated and extreme environments (example quotes:*Lack of sleep slightly increased irritability; Caught in the middle of a conflict; Irritating crew member; The end of the mission is making me feel a bit sad about it coming to an end and that I have to leave my friends and "return to Earth"*.

Among factors that could improve their state and alleviate asthenia symptoms the importance of positive interpersonal relationships and a supportive team environment emerges as a prominent theme. Participants frequently highlight the significance of good fellowship, teamwork, and emotional backing from fellow crew members. This underscores the role of social connections in fostering a sense of belonging and reducing the impact of negative experiences. A focus on engagement and task-oriented activities is also evident. Participants often mention the beneficial effects of channeling their energy and emotions into tasks and challenges, which seemingly diverts their attention from potential asthenia symptoms (if the tasks are not too exhaustive). This aligns with the concept of task immersion acting as a coping mechanism. Physical well-being is another recurrent aspect. Engaging in regular workouts and physical activities is noted multiple times, indicating a link between exercise and the alleviation of asthenia symptoms. Adequate sleep, highlighted as a potential improvement, further reinforces the importance of proper rest in combating fatigue. Interestingly, participants also note the positive influence of indulging in enjoyable experiences. Cooking and consuming tasty foods, relishing the support and companionship of friends, and the satisfaction of accomplishing missions or tasks contribute positively to their well-being (example quotes:*Other crew members and limited email contact with my best friend!; Lovely and supportive company; Teamwork and challenges; Crew cooperation and emotional support; Workout, pizza, getting work done*. Participants did not consciously identify the change in asthenia symptoms with VR experiences during the mission, but this link became more explicit in post-mission interviews.

#### 3.6.2 Post-mission interviews

Overall, the mission participants endured isolation very well. In free thought-listing, the first associations that emerged after the end of the experiment were: sadness after realizing that the shared experience was ending and the feeling of building a community with the rest of the team. The latter has been discussed notably with numerous examples of mutual positive relations. The introduction of VR experiences into the mission program was considered a good idea. The control experience - Nature VR - was described as rather boring due to the lack of action and narrative. Artistic experiences were generally well assessed. They have been described as a good break from work routines, mission experiments, and a multitude of questionnaires. However, detailed experiences' assessments varied significantly (as reflected in the assessments presented above). Cinematic VR was described as poorly prepared and "terrible". The way the film was presented posed particular problems. The participants said that they wanted to move in it like in a 3D environment, but it was impossible. However, before they accepted the limitation of the 3 degrees of freedom, they experienced unpleasant sensations related to the "movement" of the ground and the entire environment along with their movement.

The participants were enthusiastic about the Interactive artistic experience. They emphasized that what positively differentiates the interactive Nightsss from other experiences is the narrative. In this experience, they could easily pick out individual elements of the story being told. They remembered the arrival of the day as very relaxing. They also mentioned the aspect of contact with others - in an interactive experience one can try to come into contact with a virtual agent. The significance of maintaining contact with real individuals and the potential role of artistic virtual experiences in facilitating such connections were also highlighted. One of the participants noted that art, in general, is an important element that can influence the course of the mission. As an example, she mentioned making music together. This is another reference to building a community as an essential element of the mission. SCL measurements performed before and after the VR experiment had a somewhat surprising effect. They were remembered as an important moment and even an element of experience. They let the participants relax and focus on themselves for a few minutes. This could also affect the declarative measures. In summary, participants would like to use VR during similar missions, especially if it would allow them to contact virtual humans, regardless of whether they were other people in the form of avatars or virtual agents. They described artistic VR experiences as desirable.

### 4 DISCUSSION

The outcomes of our study provide compelling insights into the potential of interactive artistic virtual environments (VEs) as a promising alternative to nature-centric settings, and underscore the nuanced interplay between their effectiveness and user preference. Our findings resonate with existing research that has expounded upon the efficacy of virtual environments depicting nature in terms of relaxation and enhanced well-being [20, 44]. In our case study, the interactive artistic VE exhibited similar efficacy, evoking positive responses from participants. The allure of the artistic environment was manifest in its heightened desirability and perceived interest. Participants particularly appreciated the inclusion of narrative elements and the presence of human figures, factors that contributed to its heightened appeal. It is noteworthy, however, that the artistic VR experience was noted for its stimulating rather than relaxing nature, a distinction underscored by psychophysiological responses. This observation intersects with feedback from participants regarding potential overstimulation, a sentiment arising from the convergence of excessive responsibilities, constant data recording, and limited habitat space due to the mission's constraints. The implications of the confined environment and the continuous presence of fellow participants further compounded this issue. In contrast, the Nature VR, though initially deemed less engaging, could potentially have facilitated relaxation and mitigated nonspecific arousal through its serene and calming content. The consistent trends in self-reported well-being and emotional responses across the VR experiences hint at the potential of these interventions to influence astronauts' emotional states positively, but there is a need for immersive and engaging content to prevent potential negative emotional shifts. Although our study provides a preliminary understanding, future research should explore these effects in larger samples and over extended mission durations. In-depth qualitative analysis of mission diaries and interviews shed light on participants' subjective experiences and revealed nuanced factors influencing asthenia symptoms. The complex interplay between interpersonal dynamics, sleep patterns, and mission-related challenges emerged as significant contributors to symptom severity. The emphasis on positive interpersonal relationships, engaging activities, physical well-being, and the importance of enjoyable experiences provides valuable insights for designing interventions to alleviate asthenia symptoms. Additionally, participants' feedback on VR experiences and their desires for future interactions with virtual humans highlight the potential for VR to serve as a means of enhancing psychological well-being and social connectedness during isolated missions. The positive reception of interactive artistic VR and its narrative elements underscore its potential to foster engagement and mitigate negative emotions.

Our investigation not only contributes to the expanding domain of virtual environments within human spaceflight but also calls for a nuanced consideration of their applicability and impact. As illustrated by recent studies demonstrating the effectiveness of VEs in astronaut training and lunar scenario simulations [12, 32], the potential of VEs in addressing challenges associated with lunar ICE conditions is gaining traction. Our work further enhances this narrative by introducing artistic virtual experiences as a potential tool in this context.

As we extrapolate our findings to the broader context of long-term space missions, it is evident that further exploration is merited. The current study, while illuminating, is inherently constrained by its case study design and limited participant pool. Consequently, a more comprehensive investigation involving larger and more diverse samples is warranted to more definitively establish the scope and impact of artistic VEs in managing the psychological dynamics of isolated and extreme space environments.

In essence, our study signifies the inception of an innovative chapter in the synergy of art and science. Artistic VR not only represents a distinctive realm of experience but also presents an avenue for addressing multifaceted challenges inherent to future space missions. Through continued research and collaboration, this novel approach to psychological support in space exploration holds the potential to reshape astronaut well-being strategies and enrich the human experience beyond Earth's boundaries.

### 4.1 Limitations

Although the study was conducted in a small sample of participants in a single 2-week mission, its methodology can be replicated in other analog space habitats to verify the findings. Isolation studies, especially in the context of space, pose a challenge of having to draw conclusions from the sample of participants, equalling the number of people in the crew. To offset this limitation we use Bayesian approach to analyse single case and small group data. [10, 40]). Another consideration in such studies is that the participants had to conduct the study themselves, under the conditions of complete isolation. Therefore, they had to be trained not only in conducting such studies but also in using all of the procedures, tools, and devices and securing gathered data (e.g. charging, resetting, and backing up device data), to avoid data loss. This has proven to pose difficulties, despite available written procedures and prior experience, which shows the need for more extensive support.

### 4.2 Conclusions and future directions

In conclusion, both nature-centric and interactive artistic virtual reality environments have demonstrated the potential to enhance well-being, with the selection of artistic VR hinging on interactivity and the presence of virtual agents. Monitoring physiological responses through wearable technology offers an additional avenue for cultivating mindfulness of internal states. Ensuring the formal aspects of VR environments are thoughtfully designed is crucial for participant comfort, and a key consideration is to provide a diverse array of choices to cater to participants' individual preferences and needs.

There is great potential in creating virtual multisensory nature and art environments for people experiencing ICE conditions, either here on Earth or in the vastness of space. Such immersive multisensory virtual environments could further increase well-being and contribute to ensuring continued performance of people and teams at risk of experiencing the negative effects of isolation. This research direction may also contribute to investigating the importance of such contributions and interventions for terrestrial applications. In this case, our findings will apply to supporting not only the elderly and the disabled but also people struggling with diseases that require or cause isolation of the patient and sometimes also their relatives and caregivers. These include diseases associated with a long-term decline in immunity (e.g., during chemotherapy for certain cancers and leukemia), immobilization after various treatments and surgeries, or isolation from others as in the case of infectious diseases (for example, COVID-19). This research direction is in line with the space exploration agencies' roadmaps, that is, NASA or ESA [1], pertaining to human factors research and its terrestrial applications. We see the potential of replicating this research on a larger scale in other analog terrains or facilities that would show different characteristics of the real mission environment, like a longer isolation duration or causing the Earth-out-of-view phenomenon [25] to study the impact of artistic VR experiences in those conditions. Moreover, the terrestrial applications of the study extend beyond simulated space habitats, for example, to other ICE environments such as medical isolation (e.g pandemic or contagion outbreaks), remote natural environments, like islands or other remote locations, underwater and polar research stations, underwater and underground habitats, such as bunkers or mining operations, submarine expeditions, and also digital isolation and loneliness. Slater and Sanchez-Vives [47] discusses the wide range of VR applications across different fields, and highlights the utility of VR in enhancing well-being in different populations. We hope that this pilot research will inspire further steps into the exploration and discussion of how artistic Immersive Virtual Reality experiences impact the well-being of astronauts or other people experiencing the negative effects of isolation and other extreme conditions.

### ACKNOWLEDGMENTS

The publication is based on research carried out as part of the project "New Forms and Technologies of Narration". Project financed under the program of the Minister of Education and Science under the name "Regional Excellence Initiative" in 2019-2022, project number 023 / RID / 2018/19, financing amount: PLN 11865100. (development of artistic virtual environments)

The research leading to these results has received funding from the EEA Financial Mechanism 2014-2021 grant no. 2019/35/HS6/ 03166 (developement of the nature virtual environment).